\documentclass[12pt,preprint]{aastex}
\begin{document}

\title{THE ABSOLUTE FLUX DISTRIBUTION OF LDS749B}

\author{R.~C.\ Bohlin\altaffilmark{1}  \&  D.\ Koester\altaffilmark{2}} 
\altaffiltext{1}{Space Telescope Science Institute, 3700 San Martin Drive, 
  Baltimore, MD 21218}
\altaffiltext{2}{Institut f\"ur Theoretische Physik und Astrophysik, 
  Universit\"at Kiel, 24098 Kiel, Germany}
\email{bohlin@stsci.edu}

\begin{abstract}

Observations from the Space Telescope Imaging Spectrograph define the flux of
the DBQ4 star LDS749B from 0.12--1.0~$\mu$m with an uncertainty of $\sim$1\%
relative to the three pure hydrogen WD primary \emph{HST} standards. With
$T_\mathrm{eff}=13575~K$, $\log g=8.05$, and a trace of carbon at
$<$1$\times10^{-6}$ of solar, a He model atmosphere fits the measured STIS
fluxes within the observational noise, except in a few spectral lines with
uncertain physics of the line broadening theory. Upper limit to the  atmospheric
hydrogen and oxygen fractions by number are 1$\times10^{-7}$ and
7$\times10^{-10}$, respectively. The excellent agreement of the model flux
distribution with the observations lends confidence to the accuracy of the
modeled IR fluxes beyond the limits of the STIS spectrophotometry. The estimated
precision of $\sim$1\% in the predicted IR absolute fluxes at 30~$\mu$m should
be better than the model predictions for Vega and should be comparable to the
absolute accuracy of the three primary WD models. 

\end{abstract}


\section{Introduction}

The DBQ4 star LDS749B (WD2129+00) has long been considered for a flux standard
(e.g.,  Bohlin et~al.\ 1990). To establish the flux on the \emph{Hubble Space
Telescope} (\emph{HST}) white dwarf (WD) flux scale, STIS spectrophotometry was
obtained in 2001--2002. The virtues of LDS749B as a flux standard include an
equatorial declination and a significantly cooler flux distribution than the
33000--61000~K primary DA standards GD71, GD153, and G191B2B. Full STIS
wavelength coverage is provided from 0.115--1.02~$\mu$m, and the peak in the SED
is near 1900~\AA. At $V=14.674$ (Landolt \& Uomoto 2007), LDS749B is among the
faintest \emph{HST} standards and is suitable for use with larger ground-based
telescopes and with the more sensitive \emph{HST} instrumentation, such as the
ACS/SBC and COS. The bulk of the STIS data was obtained as part of the FASTEX
(Faint Astronomical Sources Extention) program. Finding charts appear in
Turnshek et~al.\ (1990) and in Landolt \& Uomoto (2007); but there is a large
proper motion of 0.416 and 0.034 arcsec/yr in right ascension and declination,
respectively.

The absolute flux calibration of \emph{HST} instrumentation is based on models
of three pure hydrogen WD stars GD71, GD153, and G191B2B (Bohlin 2000; Bohlin,
Dickenson, \& Calzetti 2001; Bohlin 2003). In particular, the NLTE model fluxes
produced by by the Tlusty code (Hubeny \& Lanz 1995) determine the shape of the
flux distributions using the known physics of the hydrogen atom and of stellar
atmospheres. If there are no errors in the basic physics used to determine the
stellar temperatures and gravities from the Balmer line profiles, then the
uncertainty of 3000~K for the effective temperature of G191B2B means that the
relative flux should be correct to better than 2.5\% from 0.13 to 1~$\mu$m and
to better than 1\% from 0.35 to 1~$\mu$m. 

A model that matches the observations serves as a noise free surrogate for the
observational flux distribution and provides a reliable extrapolation beyond the
limits of the observations for use as a calibration standard for \emph{JWST}, 
\emph{Spitzer}, and other IR instrumentation. Currently, the best IR absolute
flux distributions are found in a series of papers from the epic and pioneering
work of  M.~Cohen and collaborators, i.e., the Cohen-Walker-Witteborn (CWW)
network of  absolute flux standard (e.g., Cohen, Wheaton, \& Megeath 2003; Cohen
2007). The  CWW IR standard star fluxes are all ultimately based on models for
Vega and Sirius (Cohen et~al.\ 1992). More recently, Bohlin \& Gilliland (2004)
observed Vega and published fluxes on the \emph{HST}/STIS WD flux scale. A small
revision in the  STIS calibration resulted in excellent agreement of the STIS
flux distribution with  a custom made Kurucz model with $T_\mathrm{eff}=9400$~K
(Bohlin 2007), which  is the same $T_\mathrm{eff}$ used for the Cohen et~al.\
(1992) Vega model.

The model presented here for LDS749B and archived in the CALSPEC 
database\footnote{The absolute spectral energy distributions discussed in this
paper are available in digital form at
http://www.stsci.edu/hst/observatory/cdbs/calspec.html.} should have a better
precision than the Kuruzc $T_\mathrm{eff}=9400$~K model for Vega, especially
beyond $\sim$12~$\mu$m, where the Vega's dust disk becomes important (Engleke,
Price, \& Kraemer 2006). Vega is also a pole-on rapid rotator, which may also
cause IR deviations from the flux for a single temperature model. Our modeled
flux distribution for LDS749B should have an accuracy comparable to the pure
hydrogen model flux distributions for the primary WD standards GD71, GD153, and
G191B2B.

\section{The Model}

A helium model atmosphere flux distribution for LDS749B is calculated with the
LTE code of Koester (e.g., Castanheira et~al.\ 2006) for 
$T_\mathrm{eff}=13575$~K and $\log g=8.05$. At such a cool temperature, the
differences between LTE and NLTE in the continuum flux distributions should be
$<$0.1\% from the far-UV to the IR. For example for a pure hydrogen DA, the
difference between the continua of a hot 40,000~K LTE/NLTE pair of models is 1\%
between 0.1--2.5~$\mu$m. The same maximum difference at 20,000~K is only 0.3\%.
Napiwotzki (1997) did not discuss pure He models but concludes that NLTE effects
tend to become smaller with lower effective temperature. For cool DA WDs,
Koester et~al. (1998) show that the only NLTE effect that approaches 1\% is a
deeper line core of H$\alpha$. The matter densities in helium-rich white dwarfs
are significantly higher, leading to a higher ratio of collisional versus
radiative transitions between atomic levels. The larger importance of collisions
increases the tendency towards LTE occupation numbers because of the robust
Maxwell distribution of particle velocities.

The $T_\mathrm{eff}=13575$~K is higher than the  $T_\mathrm{eff}=13000$~K
published for LDS749B (alias G26$-$10) in Castanheira et~al.\ (2006), because
only UV spectra of lower precision (\emph{IUE} heritage) were used in that
analysis. Voss et~al.\ (2007) found $T_\mathrm{eff}=14440$~K with large
uncertainty, because only line profiles in the optical range were used and $\log
g$ had to be  assumed. A trace of carbon at $10^{-6}$ of the solar C/He ratio is
included, i.e. the C/He number ratio is 3.715$\times10^{-9}$. The model mass is
0.614~$M_{\sun}$ and the stellar radius is 0.01224~$R_{\sun}$, which corresponds
to a distance of 41~pc for the measured STIS flux.

The line broadening theory for the He lines combines van der Waals, Stark, and
Doppler broadening to make a Voigt profile. However, the Stark broadening uses a
simple Lorentz profile with width and shift determined from the broadening data
in Griem (1964), instead of the elaborate calculations of Beauchamp  et~al.\
(1997). The Griem method is computationally much faster, and data are available
for more lines than are calculated by Beauchamp et~al.

The fit of the higher series He lines is much improved, if the neutral-neutral
interaction is decreased in comparison to the original formalism of the
Hummer-Mihalas occupation probabilities. A similar effect was noticed by Koester
et~al.\ (2005), and our model uses the same value of the quenching parameter
that Koester et~al.\ derived ($f=0.005$). The model wavelengths are all on a vacuum
scale.

\section{STIS Spectrophotometry}

The sensitivities of the five STIS low dispersion spectrophotometric modes have
been carefully tracked since the STIS commissioning in 1997. After correcting
for changing sensitivity with time (Stys, Bohlin, \& Goudfrooij 2004) and for
charge transfer efficiency (CTE) losses for the three STIS CCD spectral modes
(Bohlin \& Goudfrooij 2003; Goudfrooij et~al.\ 2006), STIS internal
repeatability is often better than 0.5\% (Bohlin 2003). Thus, \emph{HST}/STIS
observations of LDS749B provide absolute spectrophotometry with a precision that
is superior to ground based flux measurements, which require problematic
corrections for atmospheric extinction.

Observations with a resolution $R=1000$--1500 in four STIS modes from
1150--1710~\AA\ (G140L), 1590--3170~\AA\ (G230L), 2900--5690~\AA\ (G430L), and
5300--10200~\AA\ (G750L) were obtained in 2001--2002. Earlier observations of
LDS749B in 1997 to test the time-tagged mode were unsuccessful. Two observations
in the CCD G230LB mode overlap the wavelength coverage of the MAMA G230L but are
too noisy to include in the final combined absolute flux measurement from the
other four modes. Table~1 summarizes the individual observations used for the
final combined average, along with the unused G230LB data for completeness.

Figure 1 shows the ratios of the three individual G230L and the two G230LB
observations to the model fluxes, which are normalized to the STIS flux in the
5300--5600~\AA\ range. The excellent repeatability of STIS spectrophotometry
over broad bands is illustrated and the global average ratio over the
1750--3000~\AA\ band pass is written in each panel. This ratio is unity to
within 0.3\%, even for the shorter CCD G230LB exposures despite their almost
$3\times$ higher noise level and CTE corrections. The other CCD modes G430L and
G750L also require CTE corrections. Repeatability for all the STIS spectral
modes is comparable, i.e., the global ratio deviates rarely from unity by more
than 0.6\%.

The observations in each of the four spectral modes are averaged and the four
segments are combined. This composite standard star spectrum extends from
1150--10226~\AA\ and can be obtained at
http://www.stsci.edu/hst/observatory/cdbs/calspec.html/ along with the remainder
of the \emph{HST} standard star library (Bohlin, Dickenson, \& Calzetti 2001).
This binary fits table named $lds749b\_stis\_001.fits$ has 3666 wavelength
points and seven columns. An ascii file of the flux distribution in Table~2 is
available via the electronic version of this paper. Table~2 contains the
wavelength in \AA\ and the flux in erg cm$^{-2}$ s$^{-1}$ \AA$^{-1}$ in the
first two columns, while columns~3--4 are the Poisson and systematic uncertainty
estimates in flux units, respectively. Column~5 is the FWHM of the resolution in
\AA.  The $fits$ version has two more columns than the ascii version: Column~6
is a data-quality flag, where one is good and zero may be poor quality. The
seventh column is the exposure time in seconds. The fluxes at the shortest
wavelengths below 1160~\AA\ are unreliable because of the steepness of the
sensitivity drop-off.

\section{Comparison of the Observations with the Model}

\subsection{The Continuum}

To compare the model and observations, a convenient method of removing the slope
of the spectral energy distribution (SED) is to divide both fluxes by the same
theoretical model continuum. Small differences between the observations and the
model, either in the lines or in the actual continuum, are easily illustrated in
such plots. The theoretical continuum contains only continuum opacities with an
extrapolation across the He~\textsc{i} opacity edges at 2601, 3122, and
3422~\AA, in order to avoid discontinuities. Figure~2 shows an overview of the
comparison of the STIS fluxes with the model after division of both SEDs by this
same smooth line-and-edge-free continuum. The mean continuum level of the data
between the absorption lines agrees with the model within $\sim$1\% almost
everywhere.

The most significant deviation of the data from the model is in the broad
1400--1550~\AA\ region, where each of the three spectra comprising the G140L
average have 350,000 photo-electron events in this 150~\AA\ band. The background
level is $<$0.1\% of the net signal, so that neither counting statistics nor
background subtraction error could cause the observed $\sim$1.5\% average
disparity. Of the five low dispersion modes, G140L shows the worst photometric
repeatability of individual spectra in broad bands of $\sigma\sim$0.6\%. The
three individual spectra comprising the G140L average do show occasional
2--3$\sigma$ broadband dips within their 550~\AA\ coverage region; but the
probability of such a large excursion as 1.5\% in their average is extremely
unlikely at any particular wavelength. However, the probability is much greater
that such a large excursion could occur in some 150~\AA\ band. Individual G140L
spectra of the monitoring standard GRW+70$^{\circ}$5824 often show a broad
region differing by 1--2\% from the average. The cause of such excursions could
be flat field errors, temporal instabilities in the flat field, or other
detector effects that might make the flat field inapplicable to a narrow
spectral trace.

\subsubsection{Uncertainties in $T_\mathrm{eff}$ and $\log g$}

The uncertainty in the model $T_\mathrm{eff}$ is determined by the uncertainty
in the slope of the UV flux distribution. For a constant $\log g$ model that is
cooler or hotter by 50~K and normalized to the measured 5300--5600~\AA\ flux,
there are increasing differences with the data from 1\% near 2000~\AA\ to 2\% at
the shorter wavelengths. Such a large change in the modeled continuum level in
Figure~2 (red line) is inconsistent with the STIS flux (black line). This 50~K
uncertainty of the model $T_\mathrm{eff}$ is an internal uncertainty relative to
the temperatures of the primary WD standards GD71, GD153, and G191B2B. If a
re-analysis of the Balmer lines in these primary DA standards produces a
systematic shift in the temperature scale, this shift would be reflected in a
revised $T_\mathrm{eff}$ for LDS749B that is independent of the 50~K internal
uncertainty. A 50~K temperature difference causes a $<$0.5\% flux change in the
IR longward of 1~$\mu$m.

To estimate the uncertainty in $\log g$, models are computed at the 13575~K
baseline temperature but with an increment in $\log g$. Positive and negative
increments produce nearly mirror image changes in the flux distribution. For a
decrease of 0.7 in $\log g$, the flux decreases by a nearly uniform 1.5\% below
3600~\AA\ after normalizing to unity in the 5300--5600~\AA\ range. Increasing
the $T_\mathrm{eff}$ by the full 50~K uncertainty to 13625~K can compensate for
this flux decrease below $\sim$2000~\AA. However, the +50~K increase compensates
little in the 2500--3600~\AA\ range, leaving a disparity of $\sim$1\%. Because
this 2500--3600~\AA\ range includes some of the best S/N STIS data, a 1\%
disparity establishes the uncertainty of 0.7~dex in $\log g$ as barely
compatible with the STIS flux distribution. The IR uncertainty corresponding to
this limiting case of $T_\mathrm{eff}=13625~K$ and $\log g=7.35$ is $\sim$1\%
longward of 1~$\mu$m, because the fractional percent changes in the IR from the
higher temperature and from the lower $\log g$ are both in the same direction.

\subsubsection{Interstellar Reddening}

Another source of error in the model $T_\mathrm{eff}$ is interstellar reddening.
The standard galactic reddening curve has a strong broad feature around
2200~\AA; and a tiny limit to the extinction $E(B-V)$ is set by the precise
agreement of STIS with the model in this region of Figure~2. For the upper
temperature limit of $13575+50=13625$~K, an $E(B-V)= 0.002$ brings the reddened
model into satisfactory agreement with STIS. However, for a temperature
increment of 100~K and $E(B-V)= 0.004$, the model is $\sim$1\% high at 1300~\AA\
and $\sim$1\% low at 2200~\AA. Thus, for standard galactic reddening, $E(B-V)$
must be less than 0.004, and $T_\mathrm{eff}$ is less than 13675~K. In this case
of $T_\mathrm{eff}$ and $E(B-V)$ at these allowed limits, the IR flux beyond
1~$\mu$m is still the same as for the unreddened baseline
$T_\mathrm{eff}=13575$~K within 0.5\%.

However, Bohlin (2007) presented arguments for reddening with a weak 2200~\AA\
bump for other lines of sight with tiny amounts of extinction. Reddening curves
measured in the SMC (e.g., Witt \& Gordon 2000) are missing the 2200~\AA\
feature and can cause larger uncertainty in $T_\mathrm{eff}$. Additional
evidence for extinction curves more like those in the Magellanic clouds is
presented by Clayton et~al.\ (2000) for the local warm intercloud medium, where
the reddening is low. Changes in the shape of the flux distribution after
reddening with the SMC curve of Witt and Gordon are similar to the change in
shape with $T_\mathrm{eff}$. For example, reddening a model with 
$T_\mathrm{eff}=14130$~K by SMC extinction of 0.015 is required to make an
equally unacceptable fit as for 13675~K and galactic extinction of $E(B-V)=
~0.004$. In this extreme limiting case of SMC extinction, the IR flux beyond
1~$\mu$m is still the same within $\sim$1\% as for $T_\mathrm{eff}=13575$~K and
$E(B-V)=0$.

Despite small uncertainties in the interstellar reddening and consequent
uncertainty in $T_\mathrm{eff}$, our modeling technique still predicts the
continuum IR fluxes to 1\% from 1~$\mu$m to 30~$\mu$m. Discounting the most
pathological case of SMC reddening, the worst far-IR uncertainty is from the
combined 50~K temperature and 0.7 $\log g$ uncertainties, because the changes in
the slope from the visual band normalization region into the IR due to higher
temperature and lower $\log g$ are both in the same direction. In the absence of
modeling errors or other physical complications like IR excesses from dust
rings, the measured fluxes of LDS749B relative to the three primary WDs should
be the same as predicted by the relative fluxes of the respective models to a
precision of 1\% in the IR.

\subsection{The Stellar Absorption Lines}

\subsubsection{Hydrogen}

An upper limit on the equivalent width for H$\alpha$ of $\sim$0.1~\AA\
constrains the fraction by number of hydrogen in the atmosphere of LDS749B to
$<$1$\times10^{-6}$ of helium. However, a stricter limit of $<$1$\times10^{-7}$
is provided by the weak Ly$\alpha$ line. Because interstellar absorption at
Ly$\alpha$ could be significant, zero hydrogen is consistent with the
observations and is adopted for the final best model for LDS749B. After
normalization in the $V$~band, the continuum of a model with $1\times10^{-7}$
hydrogen composition and the baseline $T_\mathrm{eff}=13575$~K and $\log g=8.05$
agrees with the zero hydrogen baseline continuum to $\sim$0.5\% from 
Ly$\alpha$ to 30~$\mu$m.

\subsubsection{Helium}

Figure 3 compares the observed He\,\textsc{i} lines with the baseline model after
correcting the model wavelengths by the radial velocity of $-$81~km s$^{-1}$
(Greenstein \& Trimble 1967). The model is smoothed with a triangular profile of
FWHM corresponding to a resolution $R=1500$ for the MAMA spectra shortward of
3065~\AA\ and to $R=1000$ for the CCD spectra longward of 3065~\AA. In general,
the model underestimates the line strengths, even for the quenching of the
neutral-neutral interactions with $f=0.005$. There is a suggestion of some
systematic asymmetry with stronger absorption in the short wavelength side of
the line profile. This asymmetry could be in the STIS line spread function
(LSF); or perhaps, a more exact treatment of the Stark line broadening theory
would reproduce the observed asymmetries.

\subsubsection{Carbon}

With a C/He ratio of $<$1$\times10^{-6}$  solar, i.e. a C/He number ratio of
3.715$\times10^{-9}$, the modeled C~\textsc{i} and C~\textsc{ii} lines reproduce
the observations within the observational noise, as shown in Figure~4. In
particular, the agreement of the modeled C~\textsc{i}(1329)/C~\textsc{ii}(1335)
line ratio with the observed ratio means that the carbon ionization ratio
corresponds to the photospheric temperature of the star. With this small amount
of carbon, the spectral clasification of LDS749B should more properly be DBQ
(Wesemael et al.\ 1993).

\subsubsection{Oxygen}

The oxygen triplet at 1302.17, 1304.87, and 1306.04~\AA\ constrains the fraction
of oxygen in the LDS749B atmosphere. This triplet absorption feature extends
over 4~\AA\ or about seven STIS pixels; but no obvious absorption feature
appears above the noise level. After binning the STIS data by seven pixels, the
rms noise in the 1300~\AA\ region is 0.8\%. The corresponding 3$\sigma$ upper
limit to the equivalent width is 0.10~\AA, which implies an upper limit to the 
atmospheric oxygen fraction by number of 7$\times10^{-10}$ of helium.

\section{Conclusion}

In the absence of any interstellar reddening, a helium model with
$T_\mathrm{eff}=13575~K\pm50$, $\log g=8.05\pm0.7$, and a trace of carbon at
$<$1$\times10^{-6}$ of solar fits the measured STIS flux distribution for
LDS749B. The noise-free, absolute flux distribution from the model after
normalization to the observed broadband visual flux is preferred for most
purposes. This normalized model SED is a high fidelity far-UV to far-IR
calibration source; and the flux distribution is available via Table~3 in the
electronic version of the \emph{Journal}. Both the observed flux distribution
and the modeled fluxes are also available from the CALSPEC
database.\footnote{http://www.stsci.edu/hst/observatory/cdbs/calspec.html/.}

\acknowledgments

Primary support for this work was provided by NASA through the Space Telescope
Science Institute, which is operated by AURA, Inc., under NASA contract
NAS5-26555. Additional support came from DOE through contract number C3691 from
the University of California/Lawrence Berkeley National Laboratory. This
research has made use of the SIMBAD database, operated at CDS, Strasbourg,
France.

\clearpage
\begin{figure} 
\centering
\includegraphics*[height=7in]{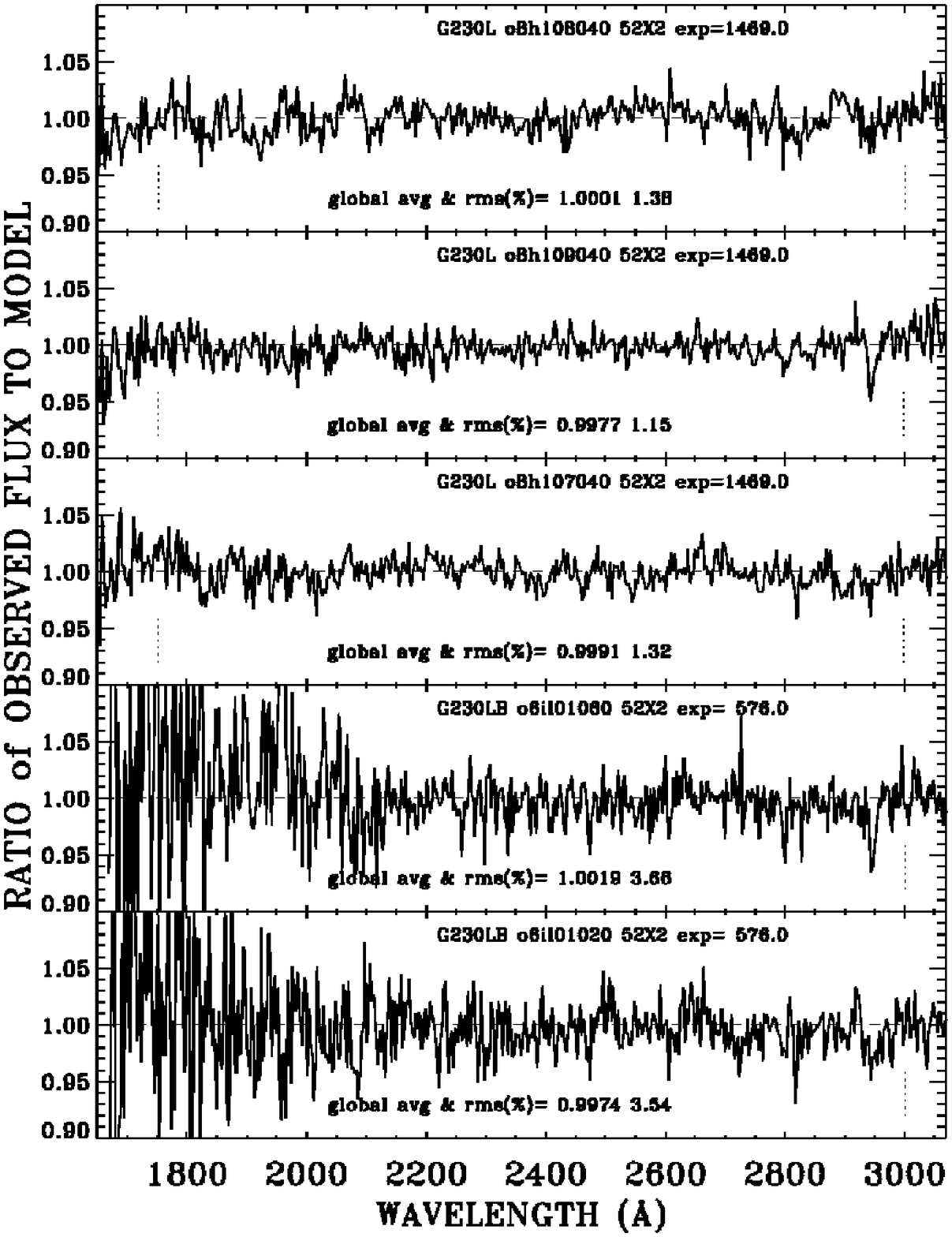}
\caption{
Ratio of individual STIS observations to the LDS749B model fluxes. Both the data
and the model are binned to the  $\sim$2~pixel resolution of STIS before
dividing. The global average and rms for the range between the vertical dotted
lines are written in each panel along with the identifying information for each
observation: spectral mode, root name, aperture, and exposure time in
seconds. The noisier, short-exposure CCD mode G230LB data shown in the bottom
two panels are not used in the final average of the observations.}
\end{figure}

\begin{figure} 
\centering
\includegraphics*[height=\textwidth,angle=90]{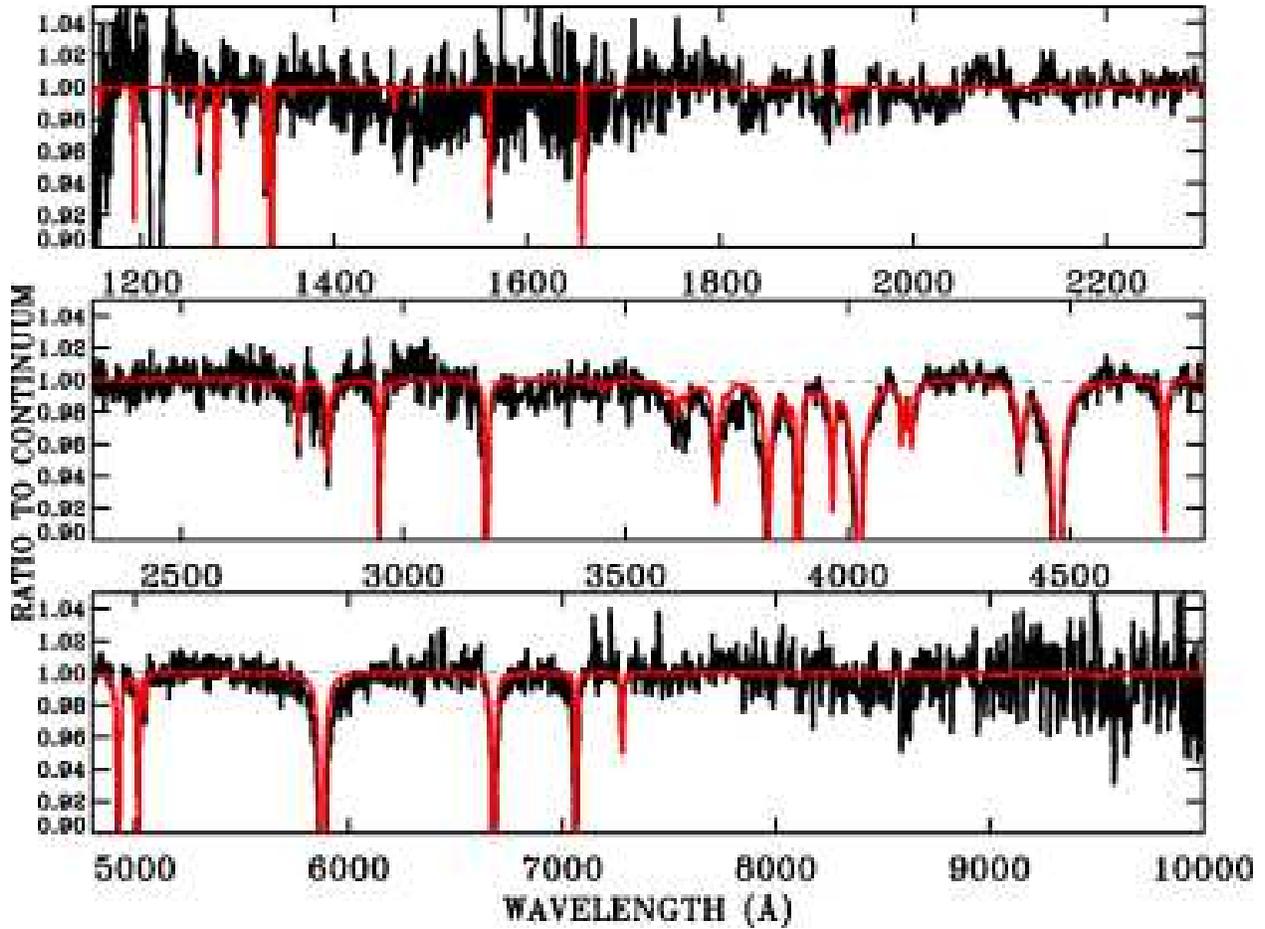}
\caption{
Comparison of the STIS observations with the model after scaling both SEDs by
the same edge-free theoretical model continuum. Both the model and its
theoretical continuum flux distributions are normalized to match the average
STIS flux in the 5300--5600~\AA\ region. The model flux is smoothed to the
approximate STIS resolution of $R=1500$ for the MAMA spectra (shortward of
3065~\AA) and to $R=1000$ for the CCD spectra (longward of 3065~\AA). A thin
dashed line marks the unity level.}
\end{figure}

\begin{figure} 
\includegraphics*[height=\textwidth,angle=90]{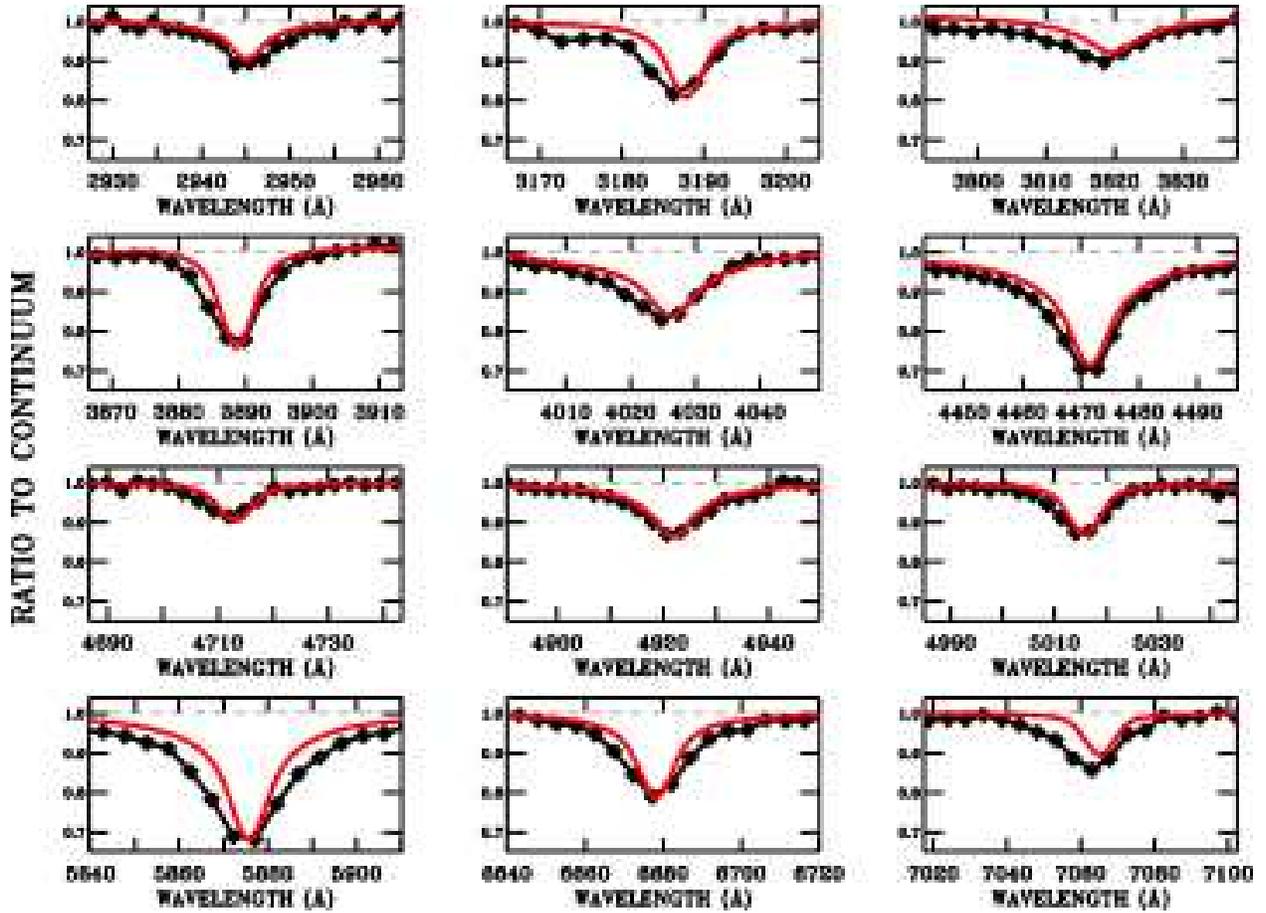}
\caption{The 12 strongest He\,\textsc{i} lines in LDS749B as in Figure 2, except
on an expanded wavelength scale.}
\end{figure}

\begin{figure} 
\includegraphics*[height=\textwidth,angle=90]{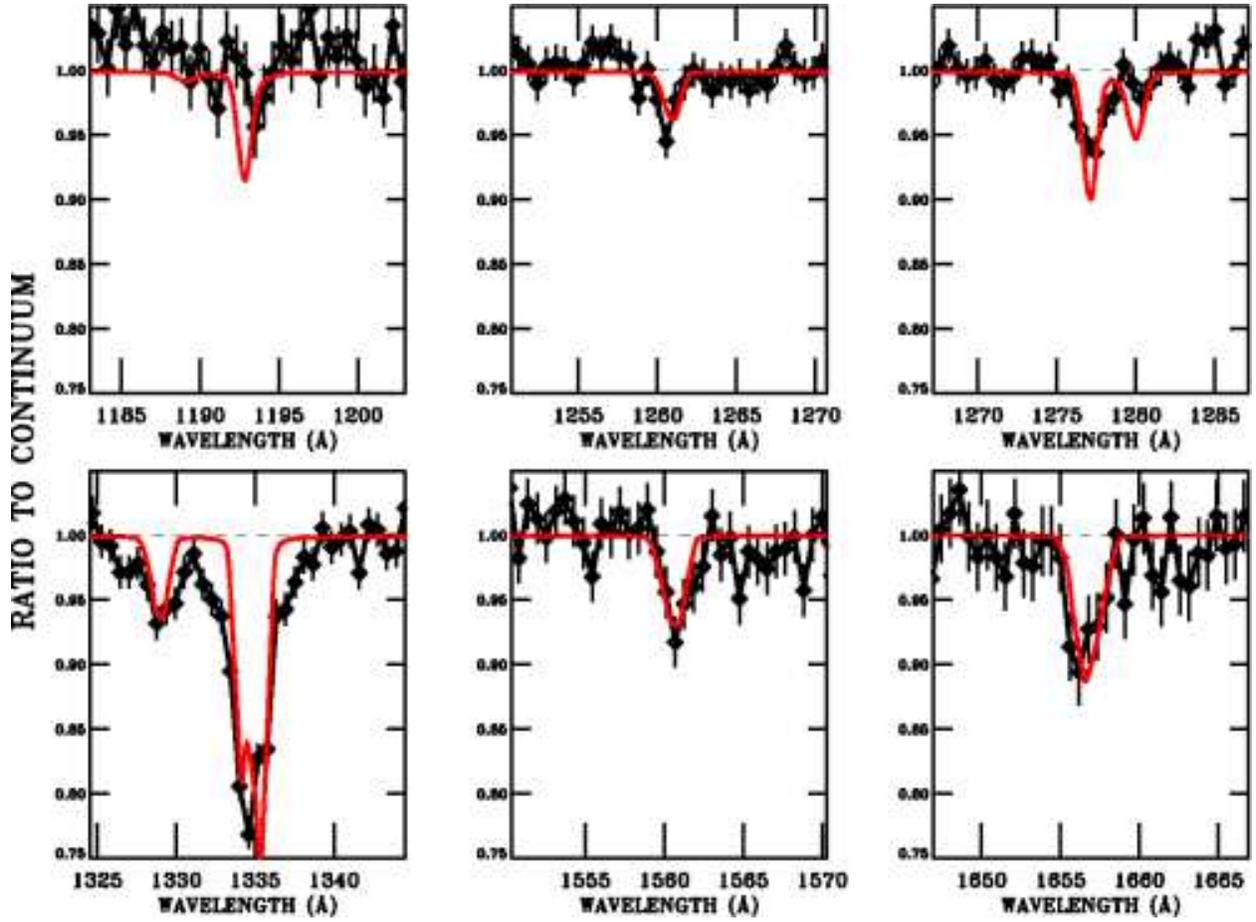}
\caption{Carbon lines in LDS749B as in Figure 2, except on an expanded wavelength
scale. The error bars are 1$\sigma$.}
\end{figure}

\clearpage
\begin{deluxetable}{lllcc}
\tablewidth{0pt}
\tablecaption{Journal of Observations}
\tablehead{\colhead{Rootname} &\colhead{Mode} &\colhead{Aperture} &\colhead{Date} 
&\colhead{Exptime(s)}}
\startdata
O6IL01020  &G230LB  &52$\times$2      &2001-07-13 &\phn576\\
O6IL01030  &G430L   &52$\times$2      &2001-07-13 &\phn619\\
O6IL01040  &G750L   &52$\times$2      &2001-07-13 &1380\\
O6IL01060  &G230LB  &52$\times$2      &2001-07-13 &\phn576\\
O6IL01070  &G430L   &52$\times$2      &2001-07-13 &\phn576\\
O6IL01080  &G750L   &52$\times$2      &2001-07-13 &1584\\
O8H107010  &G140L   &52$\times$2      &2002-09-01 &2266\\
O8H107020  &G430L   &52$\times$2      &2002-09-01 &\phn400\\
O8H107030  &G430L   &52$\times$2E1    &2002-09-01 &\phn400\\
O8H107040  &G230L   &52$\times$2      &2002-09-01 &1469\\
O8H108010  &G140L   &52$\times$2      &2002-10-16 &2266\\
O8H108020  &G430L   &52$\times$2      &2002-10-16 &\phn400\\
O8H108030  &G430L   &52$\times$2E1    &2002-10-16 &\phn400\\
O8H108040  &G230L   &52$\times$2      &2002-10-16 &1469\\
O8H109010  &G140L   &52$\times$2      &2002-12-27 &2266\\
O8H109020  &G430L   &52$\times$2      &2002-12-27 &\phn400\\
O8H109030  &G430L   &52$\times$2E1    &2002-12-27 &\phn400\\
O8H109040  &G230L   &52$\times$2      &2002-12-27 &1469
\enddata
\end{deluxetable}

\begin{deluxetable}{ccccc}
\tablewidth{0pt}
\tablecaption{Measured STIS Absolute Flux Distribution for LDS749B}
\tablehead{Wavelength &Flux &Poisson &Systematic &FWHM}
\startdata
      1150.31  &1.68860e-14  &8.09170e-16  &1.68860e-16      &1.16699\\
      1150.89  &1.59170e-14  &7.54240e-16  &1.59170e-16      &1.16687\\
      1151.47  &1.51450e-14  &7.07050e-16  &1.51450e-16      &1.16687\\
      1152.06  &1.57620e-14  &7.02910e-16  &1.57620e-16      &1.16699\\
      1152.64  &1.54940e-14  &6.71560e-16  &1.54940e-16      &1.16699\\
\enddata
\tablecomments{The complete version of this table is in the electronic
edition of the \emph{Journal}.  The printed edition contains only a sample.}
\end{deluxetable}

\begin{deluxetable}{ccc}
\tablewidth{0pt}
\tablecaption{Model Absolute Flux Distribution for LDS749B}
\tablehead{Wavelength &Flux &Continuum}
\startdata
     900.000  &8.7713311e-15  &8.7863803e-15\\
     900.250  &8.7785207e-15  &8.7950878e-15\\
     900.500  &8.7853486e-15  &8.8037809e-15\\
     901.000  &8.7975857e-15  &8.8212095e-15\\
     901.500  &8.8064389e-15  &8.8386236e-15\\
\enddata
\tablecomments{The complete version of this table is in the electronic
edition of the \emph{Journal}.  The printed edition contains only a sample.}
\end{deluxetable}

\end{document}